\documentclass[amsmath, amsfonts, showpacs, twocolumn, aps, prb]{revtex4}
\usepackage{graphicx}
\usepackage{epsfig}
\usepackage{bm}
\usepackage{euscript}
\usepackage{dcolumn}
\usepackage{amsmath}
\usepackage{amssymb}
\usepackage[varg]{txfonts}

\begin{document}

\title{Crossover in the local density of
states of mesoscopic \\ superconductor/normal-metal/superconductor
junctions}
\author{Alex Levchenko}

\affiliation{School of Physics and Astronomy, University of
Minnesota, Minneapolis, MN, 55455, USA}

\begin{abstract}
Andreev levels deplete energy states above the superconductive gap,
which leads to the peculiar nonmonotonous crossover in the local
density of states of mesoscopic
superconductor/normal-metal/superconductor junctions. This effect is
especially pronounced in the case when the normal metal bridge
length $L$ is small compared to the superconductive coherence length
$\xi$. Remarkable property of the crossover function is that it
vanishes not only at the proximity induced gap $\epsilon_g$ but also
at the superconductive gap $\Delta$. Analytical expressions for the
density of states at the both gap edges, as well as general
structure of the crossover are discussed.
\end{abstract}

\date{May 5, 2008}


\maketitle

Experimental advances in probing systems at the mesoscopic scale
\cite{Exp-1,Exp-2,Exp-3,Exp-4,Exp-5} revived interest to the
proximity related problems in superconductor -- normal metal (SN)
heterostructures.~\cite{review} The most simple physical quantity
reflecting proximity effect is the local density of states (LDOS)
$\rho(\epsilon,\mathbf{r})$, which can be measured in any spatial
point $\mathbf{r}$ at given energy $\epsilon$ using scanning
tunneling microscopy. The effects of superconductive correlations on
the spectrum of a normal metal are especially dramatic in restricted
geometries. For example, in the case of superconductor--normal
metal--superconductor (SNS) junction, proximity effect induces an
energy gap $\epsilon_g$ in excitation spectrum of a normal metal
with the square root singularity
$\rho(\epsilon,\textbf{r})\propto\sqrt{\epsilon/\epsilon_{g}-1}$ in
the density of states just above the threshold
$\epsilon-\epsilon_{g}\ll\epsilon_g$
Ref.~\onlinecite{Golubov,Belzig,Zhou,Fominov,Heikkila,Hammer} (here
and in what follows, $\rho$ will be measured in units of the bare
normal metal density of states $\nu$ at Fermi energy). The most
recent theoretical interest was devoted either to
\textit{mesoscopic}~\cite{Frahm,Altland,Vavilov,Meyer} or
\textit{quantum}~\cite{Titov,Lamacraft,OSF,Silva} fluctuation
effects on top of mean--field
results~\cite{Golubov,Belzig,Zhou,Fominov,Heikkila,Hammer} that
smear hard gap below $\epsilon_{g}$ and lead to the so called
\textit{subgap tail states} with nonvanishing
$\rho\propto\exp\big[\!-\mathrm{g}(1-\epsilon/\epsilon_{g})^{(6-d)/4}\big]$
at $\epsilon_{g}-\epsilon\lesssim\epsilon_g$, where $\mathrm{g}$ is
the dimensionless normal wire conductance and $d$ is the effective
system dimensionality. The latter is essentially a nonperturbative
result that requires instantonlike approach within
$\sigma$--model~\cite{Lamacraft,OSF} or relies on methods of random
matrix theory.~\cite{Vavilov,Titov} Surprisingly, after all of these
advances, there is something interesting to discuss about proximity
induced properties of the SNS junctions even at the level of
quasiclassical approximation by employing Usadel
equations.~\cite{Usadel} The purpose of this work is to point out a
subtle feature of the crossover in the local density of states of
mesoscopic SNS junctions. The latter was seen in some early and
recent studies,~\cite{Belzig,Heikkila,Hammer,Altland,Wilhelm}
however, neither emphasized nor theoretically addressed.

To this end, consider normal wire (N) of length $L$ and width $W$
located between two superconductive electrodes (S). In what follows,
we concentrate on diffusive quasi--one--dimensional geometry and the
short wire limit $L\ll\xi$, where $\xi=\sqrt{D_{S}/\Delta}$ is
superconductive coherence length, with $D_{S}$ as the diffusion
coefficient in the superconductor and $\Delta$ as the energy gap
(hereafter, $\hbar=1$). The center of the wire is chosen to be at
$x=0$ and boundaries with superconductors at $x=\pm L/2$
correspondingly, where $x$ is the coordinate along the wire.

\begin{figure}
\includegraphics[width=8.5cm]{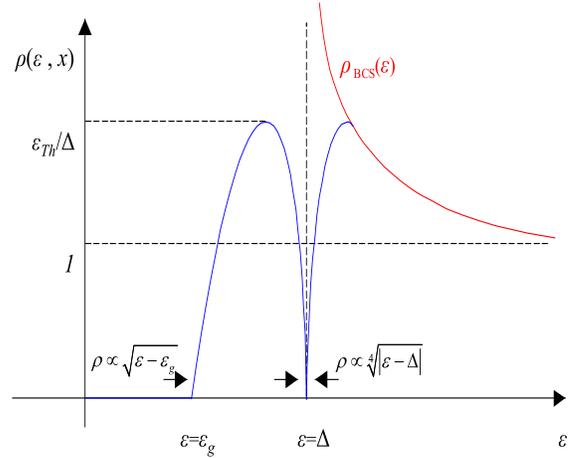}
\caption{(Color online) Schematic plot for the local density of
states crossover in the short $L\ll\xi$ diffusive SNS
junction.}\label{Fig-Crossover}
\end{figure}

Under the condition $L\ll\xi$, the proximity effect is especially
strong --- superconductive correlations penetrate the entire volume
of the normal region. As the result, the induced energy gap in
normal wire $\epsilon_g$ is large and turns out to be of the same
order as the gap in the superconductor itself
$\epsilon_g=\Delta-\delta\Delta$, with the finite size correction
$\delta\Delta\sim\Delta^{3}/\epsilon^{2}_{Th}\ll\Delta$, where
$\epsilon_{Th}=D_{N}/L^{2}$ is the Thouless energy and $D_{N}$ is
the diffusion coefficient in the normal bridge. The latter should be
contrasted to the long wire limit $L\gg\xi$, where the proximity
effect is weak and the induced gap is
$\epsilon_{g}\sim\epsilon_{Th}\ll\Delta$. Just above the gap
$\epsilon-\epsilon_g\ll\epsilon_{g}$, density of states has square
root singularity, which is similar to the $L\gg\xi$
case,~\cite{Golubov,Belzig,Zhou,Fominov,Heikkila,Hammer} see
Fig.~\ref{Fig-Crossover}, which is a robust property of
quasiclassical approximation. However, the prefactor for $L\ll\xi$
is significantly enhanced
$\rho(\epsilon,x)\propto(\epsilon_{Th}/\Delta)^{2}
\sqrt{\epsilon/\epsilon_{g}-1}$ (note here that proportionality sign
implies characteristic energy dependence, while the exact numerical
coefficient is different for a given coordinate $x$ along the wire).
The value of $\epsilon_g$ is a property of the spectrum, thus it is
$x$ independent.

One would naturally expect that above the proximity induced gap
$\epsilon_g$, local density of states goes through the maximum
$\rho_{\mathrm{max}}\!\!=\!\!\rho(\epsilon\!\!=\!\!\Delta,x)$ and
then crosses over to the BCS like DOS
$\rho_{BCS}=\epsilon/\sqrt{\epsilon^{2}-\Delta^{2}}$ at
$\epsilon>\Delta$, which finally saturates to unity $\rho\to1$ at
$\epsilon\to\infty$. Surprisingly, however, the crossover scenario
is different. The density of states indeed reaches the maximum,
which occurs at $\epsilon\sim\Delta-\delta\Delta/2$, but then
decreases and vanishes to zero at superconductive gap $\Delta$ with
quarctic power law behavior
$\rho(\epsilon,x)\propto(\epsilon_{Th}/\Delta)^{3/2}
\sqrt[4]{|\epsilon/\Delta-1|}$ for
$|\epsilon-\Delta|\lesssim\delta\Delta$. Finally, $\rho(\epsilon,x)$
grows back for $\epsilon>\Delta$ and at the energy scale
$\epsilon^{*}\!\sim\!\Delta\!+\!\delta\Delta$ crosses over to the
$\rho_{BCS}$. Indeed, observe that
$\rho_{BCS}(\epsilon^{*})\sim\epsilon_{Th}/\Delta\sim\rho(\epsilon^{*},x)$,
see Fig.~\ref{Fig-Crossover}. It is important to mention here that
the discussed feature appears only at the level of \textit{local}
density of states. Energy dependence of the \textit{global} density
of states, which is integrated over the entire volume, does not show
dip at $\epsilon=\Delta$ (this point will be discussed later in the
text).

It seems that crossover picture presented in the
Fig.~\ref{Fig-Crossover} was already numerically seen in previous
studies,~\cite{Heikkila,Hammer,Altland,Wilhelm} however the origin
of the soft gap at $\epsilon=\Delta$ was never addressed. One should
also note that numerics was performed always for the not too short
junctions $L\gtrsim\xi$. It is certainly unusual to find zero in the
density of states at the gap edge $\Delta$. In order to get some
insight into this bizarre feature, let us consider a \textit{toy
model}, which crudely mimics the system under consideration and
gives some hints for qualitative understanding. Quantitative theory
of the outlined crossover will be developed following this
discussion.

Imagine chaotic quantum dot (QD) sandwiched between two
superconductors. A quantum dot is really a zero dimensional system,
in the sense that $L/\xi\to0$ (or equivalently,
$\Delta/\epsilon_{Th}\to0$). However, one will momentarily see that
finite tunneling rate into the dot $\Gamma$ plays an effective role
of Thouless energy $\epsilon_{Th}$. By following
Ref.~\onlinecite{Beenakker} one may construct scattering states and
determine quasiparticles excitation spectrum. If in addition we
assume that QD supports only one transverse propagating mode, then
the discrete spectrum obtained from the poles of the scattering
matrix $\mathcal{S}(\epsilon)$ consists of a single nondegenerate
Andreev state at energy $\epsilon_{o}\in[0,\Delta]$, satisfying
$\Omega(\epsilon)+\Gamma\epsilon^{2}\sqrt{\Delta^{2}-\epsilon^{2}}=0\,,
$ where the function $\Omega(\epsilon)$ is defined by
$\Omega(\epsilon)=(\Delta^{2}-\epsilon^{2})(\epsilon^{2}-\Gamma^{2}/4)$.
The density of states in the superconductor--quantum
dot--superconductor system is given by
$\rho(\epsilon)=\rho_{BCS}(\epsilon)+\delta\rho(\epsilon)$, where
\begin{equation}\label{DOS-QD}
\delta\rho(\epsilon)=\frac{1}{2\pi
i}\frac{\mathrm{d}}{\mathrm{d}\epsilon}\ln\!\left(\!
\frac{\Omega(\epsilon)+i\Gamma\epsilon^{2}\sqrt{\epsilon^{2}-\Delta^{2}}}
{\Omega(\epsilon)-i\Gamma\epsilon^{2}\sqrt{\epsilon^{2}-\Delta^{2}}}\right)\,
,
\end{equation}
which follows from the relation $\rho(\epsilon)=\rho_{BCS}+(1/2\pi
i)(\mathrm{d}/\mathrm{d}\epsilon)\ln\mathrm{Det}\mathcal{S}(\epsilon)$
between DOS and the scattering matrix.~\cite{relation} It is easy to
check that in the limit $\Gamma\gg\Delta$ Andreev level is
positioned at $\epsilon_o\approx\Delta-8\Delta^{3}/\Gamma^{2}$,
which resembles the expression for $\epsilon_g$ in the case of the
diffusive wire discussed above, where $\Gamma$ indeed plays the role
of Thouless energy. The presence of an Andreev level changes density
of states $\rho(\epsilon)$ in two ways. The first contribution
$\delta\rho_{1}(\epsilon)$ originating from
$\mathrm{d}\ln\mathrm{Det}\mathcal{S}/\mathrm{d}\epsilon$ term of
Eq.~\eqref{DOS-QD} belongs to the subgap part of the spectrum
$\epsilon\in[0,\Delta]$ and has structure of the form
\begin{equation}\label{Andreev-dos-1}
\delta\rho_{1}(\epsilon)=\frac{1}{2\pi
i}\frac{\mathrm{d}}{\mathrm{d}\epsilon}\ln\left(\frac{\epsilon-\epsilon_{o}-i0}
{\epsilon-\epsilon_{o}+i0}\right)=\delta(\epsilon-\epsilon_o)\, ,
\end{equation}
which is nothing else but DOS associated with the single Andreev
state. Most interestingly, Andreev level changes $\rho(\epsilon)$
above the superconducting gap as well. Indeed, it follows from the
Eq.~\eqref{DOS-QD} that at energies $\epsilon>\Delta$ the density of
states $\rho(\epsilon)$ gets the correction
\begin{equation}\label{Andreev-dos-2}
\delta\rho_{2}(\epsilon)\!=\!-\rho_{BCS}(\epsilon)\frac{\Gamma}{\pi}
\frac{\big(2\Delta^{2}-\epsilon^{2}\big)\Gamma^{2}/4-\epsilon^{2}\big(\epsilon^{2}+\Delta^{2}\big)}
{\big(\epsilon^{2}-\Delta^{2}\big)\big(\epsilon^{2}-\Gamma^{2}/4\big)^{2}+\Gamma^{2}\epsilon^{4}}.
\end{equation}
Observe that in the immediate vicinity of the superconductive gap
$\epsilon-\Delta\lesssim\Delta^{3}/\Gamma^{2}$ the correction
$\delta\rho_{2}(\epsilon)\approx-(\Gamma/4\pi\Delta^{2})\rho_{BCS}(\epsilon)$
is \textit{negative}, implying that Andreev level
\textit{suppresses} bulk superconductive density of states
$\rho_{BCS}(\epsilon)$ above the gap. At energy
$\epsilon^{*}\sim\Delta+\Delta^{3}/\Gamma^{2}$ the correction given
by Eq.~\eqref{Andreev-dos-2} reaches its maximum, while $\rho_{BCS}$
is recovered at large energies $\epsilon\gtrsim\Gamma$, where
$\delta\rho_{2}$ decays as
$\delta\rho_{2}(\epsilon)\approx\Gamma/\pi\epsilon^{2}$. Based on
this example, it appears that Andreev level tends to
\textit{deplete} bulk BCS density of states at energies
$\epsilon-\Delta\lesssim\Delta^{3}/\Gamma^{2}$. One should note that
the Andreev level leads to pure redistribution of the energy states
--- there are no additional states above the gap; indeed,
$\int^{\infty}_{\Delta}\delta\rho_{2}(\epsilon)\mathrm{d}\epsilon\equiv0$.
If QD supports not only one but a large number of the transverse
propagating channels, then cummulative \textit{negative}
$\delta\rho_{2}(\epsilon)$ may compensate $\rho_{BCS}(\epsilon)$,
which leads to the vanishing density of states $\rho(\epsilon)$ at
the gap edge $\epsilon=\Delta$, see Fig.~\ref{Fig-Crossover}.

Having discussed the qualitative picture of the crossover, let us
now turn to the quantitative description. The quasiclassical
approach to diffusive SNS structures is based on the Usadel
equation~\cite{Usadel} for the retarded Green's function
$\EuScript{G}^{R}(\mathbf{r},\epsilon)$. For the latter, we will
employ angular parametrization~\cite{Belzig-review}
$\EuScript{G}^{R}=\tau_{z}\cos\theta+\tau_{x}\sin\theta\cos\phi
+\tau_{y}\sin\theta\sin\phi$, where $\tau_{i}$ is the set of Pauli
matrices. In the absence of the phase difference between the S
terminals, one can set $\phi\equiv0$ and Usadel equations for
quasi--one--dimension SNS geometry acquires the form
\begin{equation}\label{Usadel-eqs}
\left\{\!\!\begin{array}{ll}
D_{N}\big(\mathrm{d}^{2}\theta_{N}/\mathrm{d}x^{2}\big)+2i\epsilon\sin\theta_{N}=0
& |x|\leqslant L/2 \\
D_{S}\big(\mathrm{d}^{2}\theta_{S}/\mathrm{d}x^{2}\big)+2i\epsilon\sin\theta_{S}+
2\Delta\cos\theta_{S}=0 & |x|>L/2
\end{array}\right.
\end{equation}
where $\theta_{N(S)}(\epsilon,x)$ are the Green's function angles in
N(S) parts of the junction, correspondingly. In writing
Eq.~\eqref{Usadel-eqs}, we assumed step function pair--potential
$\Delta(x)=\Delta\eta(|x|-L/2)$, with $\eta(x)=1$ if $x>0$ and
$\eta(x)=0$ otherwise. The applicability of this approximation
relies on the condition that the width $W$ of the junction is small
compared to the coherence length. In this case, nonuniformities in
$\Delta(\mathbf{r})$ extend only over the distance of order $W$ from
the junction, which is due to the geometrically constrained
influence of the narrow junction on the bulk
superconductor.~\cite{Likharev}

In the absence of additional tunnel barriers at SN--interfaces,
Eqs.~(\ref{Usadel-eqs}a) and (\ref{Usadel-eqs}b) are supplemented by
the following boundary conditions:~\cite{KL}
\begin{subequations}\label{Boundary-Cond}
\begin{equation}
\theta_{N}(\epsilon,x)|_{x=\pm L/2}=\theta_{S}(\epsilon,x)|_{x=\pm
L/2}\, ,\quad\end{equation}
\begin{equation}
\left.\sigma_{N}\frac{\mathrm{d}\theta_{N}(\epsilon,x)}{\mathrm{d}x}\right|_{x=\pm
L/2}=\left.\sigma_{S}\frac{\mathrm{d}\theta_{S}(\epsilon,x)}{\mathrm{d}x}\right|_{x=\pm
L/2}\, ,
\end{equation}
\end{subequations}
where $\sigma_{N(S)}$ are the conductivities of normal metal and
superconductor. Knowing solutions of Usadel equations, one finds
local density of states from the general expression
\begin{equation}\label{DOS-def}
\rho(\epsilon,x)=\mathrm{Re}[\cos\theta(\epsilon,x)].
\end{equation}

For future convenience, we rotate the Green's function angles as
$\theta_{N}=\pi/2+i\vartheta_{N}$ and
$\theta_{S}=\theta_{BCS}+i\vartheta_{S}$, where
$\theta_{BCS}=\pi/2+i\mathrm{arsinh}\gamma$ with
$\gamma=\varepsilon/\sqrt{1-\varepsilon^{2}}$, and introduce
dimensionless variables $\varepsilon=\epsilon/\Delta$,
$\lambda=x/L$. After the rotation, Usadel Eq.~(\ref{Usadel-eqs}b)
for superconducting sides of the junction becomes real and can be
easily integrated, providing
\begin{equation}\label{Usadel-S-sol}
\vartheta_{S}(\varepsilon,\lambda)=4\,\mathrm{artanh}
\big\{\exp[-(|\lambda|-1/2)/\xi_{S}]\big\}\, ,
\end{equation}
where we have introduced the energy depending superconductive
coherence length $\xi^{-1}_{S}=\sqrt{\frac{2\Delta
D_{N}}{\epsilon_{Th}D_{S}}}\sqrt[4]{1-\varepsilon^{2}}$. Equation
(\ref{Usadel-eqs}a) may also be exactly solved in terms of elliptic
functions; however, this exact solution is not needed for our
purposes. Indeed, observe that it follows from the boundary
condition [Eq.~(\ref{Boundary-Cond}a)] that in the energy range
$1-\varepsilon\ll1$, the normal metal phase
$\mathrm{Re}\big[\vartheta_{N}(\varepsilon,\lambda)\big]\propto\mathrm{arcsinh}\gamma
\approx\ln\sqrt{\frac{2}{1-\varepsilon}}\gg1$ is large everywhere in
the N part of the junction. Thus, one may approximate
$\sin\theta_{N}\approx\exp(\vartheta_{N})/2$ and solve
Eq.~(\ref{Usadel-eqs}a) in closed form
\begin{equation}\label{Usadel-N-sol}
\vartheta_{N}(\varepsilon,\lambda)=\vartheta_{0}(\varepsilon)-
\ln\big[\cosh^{2}(\lambda/\xi_{N})\big]\,,
\end{equation}
with normal side coherence length being defined as
$\xi^{-1}_{N}=\sqrt{\frac{2\varepsilon\Delta}{\epsilon_{Th}}}$ and
$\vartheta_{0}=\vartheta_{N}(\varepsilon,0)$. We now introduce
$u_{0}=\exp\vartheta_{0}$ and
$u_{S}=\exp(\mathrm{arcsinh}\gamma+\vartheta^{B}_{S})$, where
$\vartheta^{B}_{S}=\vartheta_{S}(\varepsilon,\pm1/2)$, to rewrite
the boundary condition [Eq.~(\ref{Boundary-Cond}b)] as
\begin{equation}\label{Boundary-Cond-new}
u_{S}/\gamma-2=\varkappa\gamma(u_{0}-u_{S})\,,
\end{equation}
where the interface parameter
$\varkappa=\sigma^{2}_{N}D_{S}/\sigma^{2}_{S}D_{N}$ measures the
mismatch of conductivities and diffusion coefficients at the SN
boundaries. By using solutions \eqref{Usadel-S-sol} and
\eqref{Usadel-N-sol}, together with Eq.~\eqref{Boundary-Cond-new},
one eliminates the unknown $u_{0}$ and arrives to the algebraic
equation for $z=u_{S}/\gamma-2$,
\begin{eqnarray}\label{Main-Eq}
\EuScript{F}(z,\kappa)=\sqrt{\frac{\gamma\varepsilon\Delta}{8\epsilon_{Th}}}\,,
\end{eqnarray}
where the single parameter $\kappa=\varkappa\gamma^{2}$ scaling
function $\EuScript{F}(z,\kappa)$ is given by
\begin{equation}\label{F}
\EuScript{F}(z,\kappa)=\sqrt{\frac{\kappa}{z+(z+2)\kappa}}\,\mathrm{arctanh}
\sqrt{\frac{z}{z+(z+2)\kappa}}.
\end{equation}
Knowing the solution of Eq.~\eqref{Main-Eq}, one finds the density
of states at the SN interfaces as
\begin{equation}\label{Dos-def-new}
\rho(\epsilon,\pm1/2)=\frac{\gamma}{2}\,\mathrm{Im}[z(\epsilon)].
\end{equation}
At the same time, $u_{S}(\epsilon)$ together with
Eqs.~\eqref{DOS-def}--\eqref{Usadel-N-sol} provide an explicit
information about the local density of states $\rho(\epsilon,x)$ at
any position $x$ along the wire.

By looking at Eq.~\eqref{Main-Eq}, one sees that its right hand side
grows to infinity when $\varepsilon\to1$, while its left hand side
has an absolute maximum for certain value of $z$. This implies that
for all energies below some threshold $\varepsilon_{g}$,
Eq.~\eqref{Main-Eq} has the only real solution for $z$ providing
$\rho(\epsilon)\equiv0$ as it follows from Eq.~\eqref{Dos-def-new}.
The condition $\varepsilon=\varepsilon_g$ when Eq.~\eqref{Main-Eq}
has a complex solution for $z$ for the first time defines the
proximity induced energy gap $\varepsilon_g$. For energies above the
gap $\varepsilon>\varepsilon_g$, the density of states is nonzero
since $\mathrm{Im}[z]\neq0$. It turns out that Eq.~\eqref{Main-Eq}
possesses two qualitatively different solutions depending on the
value of the interface parameter $\varkappa$.

\textit{The limit of strong superconductor
$\varkappa\ll\Delta^{2}/\epsilon^{2}_{Th}$}. In this case,
$\kappa\ll1$ and $\EuScript{F}$ function determined by Eq.~\eqref{F}
has the following asymptotes:
$\EuScript{F}(z,k)\approx\sqrt{z/4\kappa}$ for $z\ll1$ and
$\EuScript{F}(z,k)\approx\sqrt{\kappa/4z}\ln(2/\kappa)$ for $z\gg1$.
It means that the only relevant $z$, which determine the maximum of
$\EuScript{F}$ are those $z\sim\kappa$. In this region,
$\EuScript{F}(z,\kappa)$ may be approximated as
$\EuScript{F}(z,\kappa)\approx\sqrt{\frac{\kappa}{z+2\kappa}}\,
\mathrm{arctanh}\sqrt{\frac{z}{z+2\kappa}}.$ As a result, the
absolute maximum $\EuScript{F}_{m}=\EuScript{F}(z=z_{m})$ occurs at
point $z_{m}\approx4.5\kappa$, which corresponds to
$\EuScript{F}_{m}\approx0.5$. Then, the gap determining condition
gives
\begin{equation}\label{Gap-1}
\EuScript{F}_{m}=\sqrt{\frac{\gamma_{g}\varepsilon_{g}\Delta}
{8\epsilon_{Th}}}\quad\Rightarrow\quad\varepsilon_g=1-\frac{1}{8}
\left(\frac{\Delta}{\epsilon_{Th}}\right)^{2},
\end{equation}
where we have used the notation $\gamma_{g}=\gamma(\varepsilon_g)$.
Just above the gap $\varepsilon-\varepsilon_g\ll\varepsilon_g$, one
can expand $\EuScript{F}$ in Taylor series around the maximum
$\EuScript{F}\approx\EuScript{F}_{m}+b(z-z_{m})^{2}$, with
$b=(1/2)(\mathrm{d}^{2}\EuScript{F}/\mathrm{d}z^{2})_{z=z_{m}}$ to
find $z\approx
z_{m}+i\sqrt{\frac{1}{b}}\sqrt{\sqrt{\frac{\gamma\varepsilon\Delta}{8\epsilon_{Th}}}-
\sqrt{\frac{\gamma_{g}\varepsilon_{g}\Delta}{8\epsilon_{Th}}}}.$ By
using now definition \eqref{Dos-def-new}, one finds for the density
of states just above the proximity induced gap at SN--interface,
\begin{equation}\label{DOS-Strong-1}
\rho(\epsilon,\pm1/2)\propto\left(\frac{\epsilon_{Th}}{\Delta}\right)^{2}
\sqrt{\frac{\epsilon-\epsilon_{g}}{\epsilon_g}}\,,\quad
\epsilon-\epsilon_g\ll\epsilon_g\,,
\end{equation}
where the numerical coefficient of the order of unity was omitted.
For the other limiting case, in the vicinity of the superconductive
gap $\varepsilon\sim1$, Eq.~\eqref{Main-Eq} is solved by $
z\approx-\pi^{2}p^{2}(1-4ip)$ with
$p=\sqrt{\frac{\epsilon_{Th}}{\Delta}}\sqrt[4]{1-\varepsilon}$,
which gives for the density of states,
\begin{equation}\label{DOS-Strong-2}
\rho(\epsilon,\pm1/2)\propto\left(\frac{\epsilon_{Th}}{\Delta}\right)^{3/2}
\sqrt[4]{\frac{|\epsilon-\Delta|}{\Delta}},\quad
|\epsilon-\Delta|\lesssim\delta\Delta.
\end{equation}
This asymptotic result holds above $\Delta$ as well. Observe that at
$\epsilon\sim\epsilon_{g}+\delta\Delta/2$, Eqs.~\eqref{DOS-Strong-1}
and \eqref{DOS-Strong-2} crossover to each other, while at
$\epsilon\sim\Delta+\delta\Delta$, Eq.~\eqref{DOS-Strong-2}
crossovers to the BCS like density of states.

\textit{The limit of weak superconductor
$\varkappa\gg\Delta^{2}/\epsilon^{2}_{Th}$}. This limiting case
corresponds to the situation when $\kappa\!\gg\!1$ and the
expression for $\EuScript{F}$ greatly simplifies
$\EuScript{F}(z,\kappa)\approx\sqrt{\frac{1}{\kappa}}\frac{\sqrt{z}}{z+2}.$
At $z_m=2$, the function $\EuScript{F}$ has the maximum
$\EuScript{F}_{m}=\sqrt{1/8\kappa_{g}}$, were
$\kappa_{g}=\varkappa\gamma^{2}_{g}$, so that the gap determining
condition is different from Eq.~\eqref{Gap-1} and reads
\begin{equation}\label{Gap-2}
\sqrt{\frac{1}{8\kappa_{g}}}=\sqrt{\frac{\gamma_{g}\varepsilon_{g}\Delta}
{8\epsilon_{Th}}}\quad\Rightarrow\quad\varepsilon_{g}=1-\frac{1}{2}
\left(\frac{\varkappa\Delta}{\epsilon_{Th}}\right)^{2/3}.
\end{equation}
To calculate the asymptotes for density of states at both gap edges,
one follows the same steps as in the previous case and finds
\begin{subequations}\label{DOS-weak-1-2}
\begin{equation}
\rho(\epsilon,\pm1/2)\propto\left(\frac{\epsilon_{Th}}{\varkappa\Delta}\right)^{2/3}
\sqrt{\frac{\epsilon-\epsilon_{g}}{\epsilon_{g}}},\quad
\epsilon-\epsilon_{g}\ll\epsilon_{g},
\end{equation}
\begin{equation}
\rho(\epsilon,\pm1/2)\propto
\sqrt{\frac{\epsilon_{Th}}{\varkappa\Delta}}\sqrt[4]{\frac{|\epsilon-\Delta|}{\Delta}},\quad
|\epsilon-\Delta|\lesssim\delta\Delta^{*},
\end{equation}
\end{subequations}
where
$\delta\Delta^{*}\sim\Delta(\varkappa\Delta/\epsilon_{Th})^{2/3}$.
Equations \eqref{DOS-Strong-1}, \eqref{DOS-Strong-2} and
\eqref{DOS-weak-1-2} complement our qualitative considerations
presented at the beginning of this paper.

At this point, let us discuss the obtained results and limits of
their applicability. (i) We have studied the energy dependence of
the local density of states for short $(L\ll\xi)$ diffusive SNS
junctions. Although $\rho(\epsilon,x)$ was analytically calculated
at SN interfaces only it turns out that its energy dependence is
generic for any $x\in[-L/2,L/2]$ and given by
Eqs.~\eqref{DOS-Strong-1}, \eqref{DOS-Strong-2} and
\eqref{DOS-weak-1-2}. The exact numerical prefactor, however, is $x$
dependent and should be determined numerically. (ii) Let us stress
that the discussed feature in the $\rho(\epsilon,x)$ at
$\epsilon\sim\Delta$ disappears at the level of global
$\langle\rho(\epsilon)\rangle$, which is integrated over the volume,
density of states. Indeed, the spatial integration brings an
additional factor of $\xi^{d}_{S}$, where $d$ is an effective
dimensionality of superconductor, which is due to the long spatial
tails of Andreev states penetrating deep inside the superconductor
[note that because of these tails, it is not correct to integrate
$\rho(\epsilon,x)$ over $x\in[-L/2,L/2]$ only]. For
quasi--one--dimensional geometry discussed here, a factor of
$\xi_{S}\propto\big(\frac{1}{1-\varepsilon}\big)^{1/4}$ gained after
$x$ integration exactly compensates the dip in $\rho(\epsilon,x)$ at
$\epsilon\sim\Delta$ [see Eqs.~\eqref{DOS-Strong-2} and
(\ref{DOS-weak-1-2}b)], leading to the finite value
$\langle\rho(\epsilon\!\!=\!\!\Delta)\rangle\sim\epsilon_{Th}/\Delta$.
For $d>1$, the density of states at superconductive gap edge
$\epsilon\sim\Delta$ should diverge as a certain power--law
$\langle\rho(\epsilon)\rangle\sim(\epsilon-\Delta)^{-\mathrm{p}}$
for $\mathrm{p}>0$. (iii) It follows from the numerical
analysis~\cite{Hammer} that presence of additional tunnel barriers
at SN interfaces alters soft gap and leads to the nonzero density of
states at the gap edge $\Delta$. (iv) Finite superconductive phase
$\phi$ imposed across the junction shifts position of the proximity
induced gap~\cite{Zhou} $\epsilon_{g}(\phi)$, such that
$\epsilon_{g}(\phi=\pi)=0$, and also changes shape of the crossover
function. However, zero in the LDOS at $\epsilon=\Delta$ and
asymptotes at both gap edges persist.

Numerous useful discussions with A.~Kamenev and L.~Glazman, which
initiated and stimulated this work are kindly acknowledged. This
research is supported by DOE Grant No. 08ER46482.



\begin{thebibliography}{99}

\bibitem{Exp-1} S.~Gu\'{e}ron, H.~Pothier, Norman O.~Birge, D.~Esteve and M.~H.~Devoret,
Phys. Rev. Lett. \textbf{77}, 3025 (1996).

\bibitem{Exp-2} M.~F.~Goffman, R.~Cron, A.~Levy Yeyati, P.~Joyez, M.~H.~Devoret,
D.~Esteve, and C.~Urbina, Phys. Rev. Lett. \textbf{85}, 170 (2000).

\bibitem{Exp-3} E.~Scheer, W.~Belzig, Y.~Naveh, M.~H.~Devoret, D.~Esteve, and
C.~Urbina, Phys. Rev. Lett. \textbf{86}, 284 (2001).

\bibitem{Exp-4} A.~Anthore, H.~Pothier, and D.~Esteve, Phys. Rev.
Lett. \textbf{90}, 127001 (2003).

\bibitem{Exp-5} Yong-Joo Doh, Jorden A.~van Dam, Aarnoud L.~Roest, Erik P.~A.~M.~Bakkers,
Leo P.~Kouwenhoven, Silvano De Franceschi, Science \textbf{309}, 272
(2005).

\bibitem{review} See, e.g., \textit{Mesoscopic
Superconductivity}, edited by P.~F.~Bagwell special issue of
Superlattices Microstruct. \textbf{25}, 5--6 (1999).

\bibitem{Golubov} A.~A.~Golubov, E.~P.~Houwman, J.~G.~Gijsbertsen, V.~M.~Krasnov,
J.~Flokstra, H.~Rogalla, and M.~Yu.~Kupriyanov, Phys. Rev. B
\textbf{51}, 1073 (1995).

\bibitem{Belzig} W.~Belzig, C.~Bruder, and G.~Sch\"{o}n, Phys. Rev. B
\textbf{54}, 9443 (1996).

\bibitem{Zhou} F.~Zhou, P.~Charlat, B.~Spivak, and B.~Pannetier, J. Low Temp. Phys.
\textbf{110}, 841 (1998).

\bibitem{Fominov} Ya.~V.~Fominov and M.~V.~Feigel'man, Phys. Rev. B
\textbf{63}, 094518 (2001).

\bibitem{Heikkila} T.~T.~Heikkil\"{a}, J.~S\"{a}rkk\"{a}, and
F.~K.~Wilhelm, Phys. Rev. B \textbf{66}, 184513 (2002).

\bibitem{Hammer} J.~C.~Hammer, J.~C.~Cuevas, F.~S.~Bergeret, and W.~Belzig,
Phys. Rev. B \textbf{76}, 064514 (2007).

\bibitem{Frahm} K.~M.~Frahm, P.~W.~Brouwer, J.~A.~Melsen, and C.~W.~J.~Beenakker,
Phys. Rev. Lett. \textbf{76}, 2981 (1996).

\bibitem{Altland} A.~Altland, B.~D.~Simons and D.~Taras--Semchuk,
Adv. Phys. \textbf{49}, 321 (2000).

\bibitem{Wilhelm} F.~K.~Wilhelm and A.~A.~Golubov, Phys. Rev. B \textbf{62},
5353 (2000).

\bibitem{Vavilov} M.~G.~Vavilov, P.~W.~Brouwer, V.~Ambegaokar, and
C.~W.~J.~Beenakker, Phys. Rev. Lett. \textbf{86}, 874 (2001).

\bibitem{Meyer} J.~S.~Meyer and B.~D.~Simons, Phys. Rev. B
\textbf{64}, 134516 (2001).

\bibitem{Titov} M.~Titov, N.~A.~Mortensen, H.~Schomerus, and C.~W.~J.~Beenakker,
Phys. Rev. B \textbf{64}, 134206 (2001).

\bibitem{Lamacraft} A.~Lamacraft and B.~D.~Simons, Phys. Rev. Lett.
\textbf{85}, 4783 (2000).

\bibitem{OSF} P.~M.~Ostrovsky, M.~A.~Skvortsov, and M.~V.~Feigel'man,
Phys. Rev. Lett. \textbf{87}, 027002 (2001).

\bibitem{Silva} A.~Silva, Phys. Rev. B \textbf{72}, 224505 (2005).

\bibitem{Usadel} K.~D.~Usadel, Phys. Rev. Lett. \textbf{25}, 507
(1970).

\bibitem{Beenakker} C.~W.~J.~Beenakker and H.~van Houten,
\textit{Single-Electron Tunneling and Mesoscopic Devices}, edited by
H.~Koch and H.~L\"{u}bbig, Springer, Berlin, (1992), pp.~175-179.

\bibitem{relation} E.~Akkermans, A.~Auerbach, J.~E.~Avron, and B.~Shapiro,
Phys. Rev. Lett. \textbf{66}, 76 (1991).

\bibitem{Belzig-review} W.~Belzig, F.~K.~Wilhelm, C.~Bruder, G.~Sch\"{o}n, and
A.~D.~Zaikin, Superlattices Microstruct. \textbf{25}, 1251 (1999).

\bibitem{Likharev} K.~K.~Likharev, Rev. Mod. Phys. \textbf{51}, 101
(1979).

\bibitem{KL} M.~Yu.~Kuprianov and V.~F.~Lukichev, Sov. Phys. JETP
\textbf{67}, 1163 (1988).

\end{thebibliography}
\end{document}